\documentclass[aps,twocolumn,nofootinbib,showpacs, prd]{revtex4-2}

\usepackage{hyperref}
\usepackage{amsmath,amsfonts,amssymb}
\hypersetup{dvips,dvipdfm,colorlinks=true,urlcolor=magenta,filecolor=magenta,linktoc=page,citecolor=red,linkcolor=blue,bookmarks=true}
\usepackage{graphicx,epstopdf}
\usepackage{multirow}
\usepackage{natbib}
\begin{document}
	\title{Raychaudhuri Equation from Lagrangian and Hamiltonian formulation : A Quantum Aspect}
	\author{Madhukrishna Chakraborty} \email{chakmadhu1997@gmail.com}
	\author{Subenoy Chakraborty}\email{schakraborty.math@gmail.com (corresponding author)}
	\affiliation{Department of Mathematics, Jadavpur University, Kol - 700032, India}
	\begin{abstract}
		The paper deals with a suitable transformation related to the metric scalar of the hyper-surface so that the Raychaudhuri Equation (RE) can be written as a second order nonlinear differential equation. A first integral of this second order differential equation gives a possible analytic solution of the RE. Also, it is shown that construction of a Lagrangian (and hence a Hamiltonian) is possible, from which the RE can be derived. Wheeler-Dewitt equation has been formulated in canonical quantization scheme  and norm of it's solution (wave function of the universe) is shown to affect the singularity analysis in the quantum regime for any spatially homogeneous and isotropic cosmology. Finally Bohmian trajectories are formulated with causal interpretation and these quantum trajectories unlike classical geodesics obliterate the initial big-bang singularity when the quantum potential is included.
	\end{abstract}
	\maketitle
	\small ~Keywords :  Raychaudhuri Equation ;  Quantization; Bohmian Trajectories .
	\section{Introduction}
	The appearance of singularity in Einstein's general theory of relativity is a well-known fact. So the natural question that arises ``Is this singularity inevitable in general relativity?". Raychaudhuri in the early 1950's tried to address this issue by formulating an evolution equation for expansion scalar---the Raychaudhuri equation (RE) \cite{Raychaudhuri:1953yv},\cite{Dadhich:2007pi},\cite{Dadhich:2005qr},\cite{Ehlers:2006aa},\cite{Kar:2008zz},\cite{Kar:2006ms}. Subsequently, Hawking and Penrose formulated the seminal singularity theorems \cite{Penrose:1964wq},\cite{Hawking:1970zqf},\cite{Hawking:1973uf} in General Relativity (GR) using this RE as the main ingredient through the notion of geodesic focusing. At the singularity, there is no structure of space-time and physical laws break down. It is generally speculated that quantum effects which become dominant in strong gravity regime may alleviate the singularity problem at the classical level. In particular a quantum version of the Raychaudhuri equation may probably be useful in the context of identifying the existence of a singularity in the quantum regime. Though there is no universally accepted theory of quantum gravity \cite{Thiemann:2007pyv}, there are at present two major approaches for formulating a quantum theory of gravity-- canonical quantization \cite{DeWitt:1967yk} and path integral formulation \cite{Hawking:1978jz}. In canonical quantization , the operator version of the Hamiltonian constraint (known as Wheeler-Dewitt (WD) equation) is a second order hyperbolic functional differential equation and its solution is known as the wave function of the universe \cite{Hartle:1983ai}. However, even in simple minisuperspace models it is hard to find a solution of the WD equation. Also there is an ambiguity in operator ordering and how to know the initial conditions of the universe to have a well defined wave function. However an important feature of the Hamiltonian in the operator version is that it admits a self adjoint extension in a general sense. As a result, the conservation of probability is ensured. On the other hand, the path integral formulation is more favourable due to some definite proposals for the sum over histories (namely by Hartle \cite{Hartle:2022ykc}, Hawking \cite{Hawking:1978jz} and by Vilenkin \cite{Vilenkin:1988yd}).\\
	~~~~~~~~~~~~~~ RE may be considered as a key ingredient for the classical singularity theorems. So a general speculation about the behavior of the singularity at the quantum domain may be revealed by examining the RE in quantum settings. In this context it is worthy to mention that quantum version of the RE by Das \cite{Das:2013oda} does not allow focusing of geodesics as long as there is non zero quantum potential in Bohmian trajectories. A general view about the avoidance of singularity is the existence of repulsive terms due to quantum effects. For better clarity one can go through the works \cite{Das:2013oda}-\cite{Blanchette:2020kkk}\\
	$~~~~~~~~~~$ The quantization scheme of RE has some issues. As RE is essentially an identity in the Riemannian geometry so naturally RE cannot be obtained from a variational principle as the equation of motion for a geodesic congruence. However when we express the curvature scalar in terms of Einstein's field equations (or modified field equations) then RE is no longer an identity, rather there is some meaning as Lagrangian formulation.\\
	In the present work,  the RE has been transformed to a second order differential equation in a general space-time and it is possible to have a first integral of this second order differential equation. This first integral gives a possible solution of the RE. A general form of the Lagrangian and Hamiltonian has been formulated so that the RE can be derived from them in any dimension. Finally, a canonical quantization scheme has been presented.\\ The plan of the paper is as follows: Section II shows a possible solution of the RE, Section III deals with Lagrangian and Hamiltonian formulation to obtain RE. The canonical quantization scheme with construction of WD equation has been presented in section IV and solution of the WD equation plays an important role in identifying/avoiding singularity in spatially homogeneous and isotropic cosmological models. Section V formulates the Bohmian trajectories for the present quantum system with causal interpretation and conditions under which the trajectories can avoid the big-bang singularity are discussed. The paper ends with a brief discussion and conclusion in Section VI.
	\section{The Raychaudhuri equation and its solution}
	Let us consider a congruence of time-like geodesics in an $(n+1)$-dimensional spacetime $\mathcal{M}$. Suppose $\Sigma$ be an $n$-dimensional hyper-surface (space-like) such that the congruence of time-like geodesics are orthogonal to this hyper-surface $\Sigma$ i.e the unit velocity vector $u^{\mu}$ to the congruence is orthogonal to the hyper-surface. Let $q_{ab}$ be the induced metric on the $n$ dimensional hyper-surface $\Sigma$. If the congruence of geodesics is treated as dynamical system \cite{Alsaleh:2017ozf} then it is convenient to consider
	\begin{equation}\label{eq1}
		x(\tau)=\sqrt{q}
	\end{equation}
	as the dynamical degree of freedom ($\tau$ is the proper time and $q$=det$(q_{ab}))$. Essentially $x$ is related to the volume of the hyper-surface and $x=0$ identifies the singularity. Using the definition of the expansion scalar $\theta$ of the congruence as
	\begin{equation}
		\dfrac{d}{d\tau}\ln\sqrt{q}=\theta=\nabla_{\mu}u^{\mu}
	\end{equation}
	the evolution of $x$ takes the form,
	\begin{equation}\label{eq4}
		\dfrac{dx}{d\tau}=x\theta
	\end{equation}
	Due to orthogonality of each member of the congruence to the hyper-surface, the rotation tensor $\omega_{ab}=\dfrac{1}{2}\left(\nabla_{b}u_{a}-\nabla_{a}u_{b}\right)=0$ as a consequence of Frobenius theorem of differential geometry. So, the Raychaudhuri equation simplifies to
	\begin{equation}
		\dfrac{d\theta}{d\tau}=-\dfrac{\theta^{2}}{n}-2\sigma^{2}-\tilde{R}\label{eq4*}
	\end{equation}
	where $2\sigma^{2}=\sigma_{ab}\sigma^{ab}$, $\sigma_{ab}=\dfrac{1}{2}\left(\nabla_{a}u_{b}+\nabla_{b}u_{a}\right)-\dfrac{1}{n}\theta q_{ab}$ is the shear tensor and $\tilde{R}=R_{ab}u^a u^b$ is an effective scalar curvature, also known as the Raychaudhuri scalar. Using the evolution equation (\ref{eq4}) for the dynamical degree one may write the above Raychaudhuri equation as a second order nonlinear differential equation as
	\begin{equation}
		x\dfrac{d^{2}x}{d\tau^{2}}+\left(\dfrac{dx}{d\tau}\right)^{2}\left(\dfrac{1}{n}-1\right)+(2\sigma^{2}+\tilde{R})x^{2}=0\label{eq5}
	\end{equation}
	which has a first integral
	\begin{equation}
		\left(\dfrac{dx}{d\tau}\right)^{2}=z_{0}x^{2(1-\frac{1}{n})}-2x^{2(1-\frac{1}{n})}\int{x^{(\frac{2}{n}-1)}}(2\sigma^{2}+\tilde{R})dx
	\end{equation} with $z_{0}$, a constant of integration. So using (\ref{eq4}), we have the solution of the Raychaudhuri equation as 
	\begin{equation}
		\theta^{2}=z_{0}x^{-\frac{2}{n}}-2x^{-\frac{2}{n}}\int{x^{(\frac{2}{n}-1)}}(2\sigma^{2}+\tilde{R})dx
	\end{equation}
	One may note that unlike the field equations, the Raychaudhuri equation has nothing to do with any gravity theory. It is purely a geometric identity, but the role of gravity comes into picture through the Ricci tensor ($R_{ab}$) projected along the geodesics. Therefore, one may find the explicit expression for $\theta$ for a particular gravity theory in a four dimensional spacetime.
	\section{Lagrangian and Hamiltonian Formulation of Raychaudhuri equation}
	In the previous section, it is found that the Raychaudhuri equation can be expressed as a second order differential equation. So, it is a natural search for a Lagrangian corresponding to which the Euler-Lagrange equation gives (\ref{eq5}). According to Helmholtz \cite{Davis:1928},\cite{Davis:1929},\cite{Douglas:1941},\cite{Casetta:1941},\cite{Crampin:2010},\cite{Nigam:2016}, for a system of 'r' second order differential equations of the form
	\begin{equation}
		\mu_{\alpha}(\tau,y_{\delta},\dot{y_{\delta}},\ddot{y_{\delta}})=0 ,~~ \alpha,\delta=1,2,...,r
	\end{equation}
	( `.' indicates differentiation w.r.t proper time $\tau$), the necessary and sufficient conditions for being the Euler-Lagrange equations corresponding to a Lagrangian $L(\tau,y_{\delta},\dot{y_{\delta}})$, termed as Helmholtz conditions \cite{Davis:1928}-\cite{Nigam:2016} are given by
	\begin{eqnarray}
		\dfrac{\partial\mu_{\alpha}}{\partial\ddot{y_{\delta}}}=\dfrac{\partial\mu_{\delta}}{\partial\ddot{y_{\alpha}}}\\
		~~~~~~~~~	\dfrac{\partial\mu_{\alpha}}{\partial y_{\delta}}-\dfrac{\partial\mu_{\delta}}{\partial y_{\alpha}}=~\dfrac{1}{2}~\dfrac{d}{d\tau}\left(\dfrac{\partial\mu_{\alpha}}{\partial{\dot{y_{\delta}}}}-\dfrac{\partial{\mu_{\delta}}}{\partial{\dot{y_{\alpha}}}}\right)
	\end{eqnarray} and
	\begin{equation}
		~~~~~~~~~~~~~~~	\dfrac{\partial\mu_{\alpha}}{\partial{\dot{y_{\delta}}}}+\dfrac{\partial{\mu_{\delta}}}{\partial{\dot{y_{\alpha}}}}=2~\dfrac{d}{d\tau}~\left(\dfrac{\partial\mu_{\delta}}{\partial\ddot{y_{\alpha}}}\right)
	\end{equation} with $(\alpha,\delta)=1,2,...,r$.
	In the present context, we have a single second order differential equation (\ref{eq5}) so the above conditions reduce to
	\begin{equation}\label{eq13}
		\dfrac{d\mu}{d\dot{x}}=\dfrac{d}{d\tau}\left(\dfrac{d\mu}{d\ddot{x}}\right)
	\end{equation} with,
	\begin{equation}\label{eq14} \mu(\tau,x,\dot{x},\ddot{x})=x\ddot{x}+\left(\frac{1}{n}-1\right)\dot{x}^{2}+(2\sigma^{2}+\tilde{R})x^{2}
	\end{equation}
	A simple algebraic calculation shows that equation (\ref{eq13}) is satisfied for $\mu$ given in equation (\ref{eq14}) only for $n=\frac{2}{3}$, which is not possible as $n$ is the dimension of the hyper-surface. (If one chooses,  $\mu=\dfrac{\ddot{x}}{x}+\left(\frac{1}{n}-1\right)\dfrac{\dot{x}^{2}}{x^{2}}+(2\sigma^{2}+\tilde{R})$ then (\ref{eq13}) implies $n=2$). Thus for general '$n$', ~ (\ref{eq13}) will be satisfied for $\tilde{\mu}$ , provided
	$\tilde{\mu}=x^{\alpha}\mu$ with $\alpha=\left(\dfrac{2}{n}-3\right).$
	Therefore, one has
	\begin{eqnarray}
		\tilde{\mu}=x^{2\left(\frac{1}{n}-1\right)}\ddot{x}+\left(\frac{1}{n}-1\right)x^{\left(\frac{2}{n}-3\right)}\dot{x}^{2}+(2\sigma^{2}+\tilde{R})x^{\left(\frac{2}{n}-1\right)}\nonumber\\
		=\dfrac{d}{d\tau}\left[x^{2(\frac{1}{n}-1)}\dot{x}\right]-\left(\frac{1}{n}-1\right)x^{(\frac{2}{n}-3)}\dot{x}^{2}+h(x)x^{(\frac{2}{n}-1)}
	\end{eqnarray} provided $(2\sigma^{2}+\tilde{R})$ is  a function of $x$ alone, say $h(x)$.
	Thus one may construct the Lagrangian as
	\begin{equation}
		\mathcal{L}=\dfrac{1}{2}x^{2\left(\frac{1}{n}-1\right)}\dot{x}^{2}-V[x]\label{eq15}
	\end{equation} with,
	\begin{equation}
		\dfrac{\delta V[x]}{\delta x}=x^{(\frac{2}{n}-1)}h(x)\label{eq16}
	\end{equation}
	Now the momentum conjugate to the variable '$x$' is
	\begin{equation}\label{eq18}
		\Pi_{x}=\dfrac{\partial\mathcal{L}}{\partial\dot{x}}=x^{2(\frac{1}{n}-1)}\dot{x}
	\end{equation}
	At this point one may check that the Euler Lagrange equation corresponding to the Lagrangian in eq. (\ref{eq15}) gives back the RE in eq. (\ref{eq5}).
	So, the Hamiltonian of the system is given by
	\begin{equation}
		\mathcal{H}=\dfrac{1}{2}x^{-2\left(\frac{1}{n}-1\right)}\Pi_{x}^{2}+V[x]
	\end{equation}
	One may note that one of the Hamilton's equation of motion gives the Raychaudhuri equation (\ref{eq5}) while the other one yields the definition of momentum (\ref{eq18}).
	From the above formulation of $\tilde{\mu}$, it is clear that $\tilde{\mu}$ satisfies all the Helmholtz conditions provided $2\sigma^{2}+\tilde{R}$ is a sole function of $x$. Now we recall the RE in eq. (\ref{eq4*}) and denote the scalar $2\sigma^2+\tilde{R}$ by $R_c$. Therefore eq.(\ref{eq4*}) can be written as
	\begin{equation}
		\dfrac{d\theta}{d\tau}+\dfrac{\theta^2}{n}=-R_c
	\end{equation}	
	If $R_c>0$ then $\dfrac{d\theta}{d\tau}+\dfrac{\theta^2}{n}<0$. Integrating this inequality w.r.t $\tau$ we get,
	\begin{equation}
		\dfrac{1}{\theta(\tau)}>\dfrac{1}{\theta_0}+\dfrac{\tau}{n}.
	\end{equation}
	This shows that an initially converging hyper-surface orthogonal congruence of time-like geodesics must continue to converge within a finite value of the proper time $\tau<n\theta_0^{-1}$ thereby forming caustics. This leads to crossing/focusing of geodesics. Although the singularity here is the congruence singularity and may not be a space-time
	singularity. In most of the situations these caustics do
	not lead to any spacetime singularities, but under certain circumstances they do, leading to formation of black hole or cosmological singularities. Removal of these curvature singularities has remained a puzzle for decades. This vital consequence of the RE ultimately played a key role in the proof of the seminal singularity theorems furnished by Hawking and Penrose. The condition $R_c>0$ is called the Convergence Condition (CC) and hence we name the scalar $R_c$ as the Convergence scalar. In this context $R_c=h(x)$. L.H.S of eq. (\ref{eq16}) is the gradient of the potential corresponding to the dynamical system representing the congruence and has to be constructed using gravitational field equations. Further one finds that $R_c>(<0)$ implies force is attractive (repulsive) in nature. This hints that convergence will occur (i.e. $R_c>0$) if the matter is attractive. This is the reason why RE is regarded as the fundamental equation of gravitational attraction. \\In quantum correction of the RE, some extra terms are added with ($-R_{c}$) in the R.H.S of the classical RE so that they act as repulsive force to prevent focusing. Appearance of singularity implies convergence/focusing, hence if one can prevent focusing using this quantum correction of RE avoidance of singularity might be guaranteed. (for ref see \cite{Das:2013oda}-\cite{Blanchette:2020kkk}).
	\section{Wheeler-Dewitt equation: The canonical approach}
	To formulate the operator version of the Hamiltonian from the classical one as obtained in the previous section we carry out the canonical quantization of the system under consideration where $x$ and $\Pi_{x}$ are promoted to operators so that $[\hat x, \hat \Pi_{x}]=i\hbar$. These operators act on the geometric flow state $\Psi[x,\tau]$. In $x$- representation we have $\hat x=x,~\hat \Pi_{x}=-i\hbar\dfrac{\partial}{\partial x}$. Thus the Hamiltonian in terms of operators is given by $\tilde{H}=-\dfrac{\hbar^{2}}{2}x^{2(1-\frac{1}{n})}\dfrac{\partial^{2}}{\partial x^{2}}+V[x]$. The operator version of the Hamiltonian is given by $\tilde{H}=-\dfrac{\hbar^{2}}{2}x^{2(1-\frac{1}{n})}\dfrac{\partial^{2}}{\partial x^{2}}+V[x]$. Consequently the evolution equation of the physical state $\Psi$ can be described by the Schrodinger equation $\tilde{H}\Psi=i\hbar\dfrac{\partial}{\partial \lambda}\Psi$. One may consider this as the quantized version of the evolution of a time-like geodesic congruence having classical analogue as the Raychaudhuri equation but this is applicable to only a limited class of geometries. However in the context of cosmology, there is notion of Hamiltonian constraint and operator version of it acting on the wave function of the universe $\Psi$ i.e, $\tilde{H}\Psi=0$, known as the Wheeler Dewitt (WD) equation. This is because $\tilde{H}$ generates infinitesimal gauge transformations and since physical
	states should be invariant under gauge transformations, they should be invariant under
	the action of the group member associated to $\tilde{H}$ (its exponential). Alternatively it is
	because classical $H = 0$ on the constraint surface and hence quantum mechanically
	$\tilde{H}\Psi=0$. In the quantization process due to Dirac, the physical states or physical quantum states of the associated Hilbert Space must be annihilated by the operator version of $\mathcal{H}$. In this case the WD equation takes the form
	\begin{equation}
		\dfrac{d^{2}\Psi}{dx^{2}}-\dfrac{2}{\hbar^{2}}x^{2(\frac{1}{n}-1)}V(x)\Psi(x)=0\label{eq21}
	\end{equation}
	($\because$ for spatially homogeneous and isotropic cosmological models $V[x]=V(x)$ and due to operator ordering $\Pi_{x}^{2}$ is replaced by $[-\frac{d^{2}}{dx^{2}}-\frac{p}{x}\frac{d}{dx}]$, with $p=0$ as the ordering factor). However for spatially anisotropic models there is a problem of non unitary evolution which can be resolved by the proper choice of operator ordering in the first term of the Hamiltonian. The operator form $\tilde{H}=-\dfrac{\hbar^{2}}{2}x^{2(1-\frac{1}{n})}\dfrac{\partial^{2}}{\partial x^{2}}+V[x]$ is symmetric with norm $|\Psi|^{2}=\int_{0}^{\infty}dx~x^{2(\frac{1}{n}-1)}\Psi^{*}\Psi$ but fails to be self adjoint . However a self adjoint extension is guaranteed by Friedrichs \cite{Pal:2016ysz}. For example if we consider the operator ordering as $\tilde{H}=-\dfrac{\hbar^{2}}{2}x^{(1-\frac{1}{n})}\dfrac{\partial}{\partial x}x^{(1-\frac{1}{n})}\dfrac{\partial}{\partial x}+V[x]$ and a change in minisuperspace variable as $u=nx^{\frac{1}{n}}$ then the WD becomes
	\begin{equation}
		\left[\dfrac{-\hbar^{2}}{2}\dfrac{d^{2}}{du^{2}}+V[u]\right]\Psi(u)=0,\label{eq22}
	\end{equation}
	with symmetric norm as
	\begin{equation}
		|\Psi|^{2}=\int_{0}^{\infty}du~\Psi^{*}\Psi,
	\end{equation} provided the integral exists and finite.
	Hence the Hamiltonian admits a self adjoint extension resolving the non unitary evolution of the geodesic congruence. The WD equation (\ref{eq22}) can be interpreted as time-independent Schrödinger equation of a point particle of unit mass moving along $v$ direction in a potential field $V(v)$ and it has zero eigen  value of the Hamiltonian and the wave function of the universe is identified as the energy eigen function.  Therefore for the sake of singularity analysis in the quantum regime in case of homogeneous and isotropic/ anisotropic cosmological models (a larger class of models) we have invoked the WDW equation and inclined to find its solution in those models.
	The takeaway from the above formulation is listed below:
	\begin{enumerate}
		\item The WD equation (\ref{eq22}) can be interpreted as time-independent Schrödinger equation of a point particle of unit mass moving along $u$ direction in a potential field $V[u]$ and it has zero eigen  value of the Hamiltonian and the wave function of the universe is identified as the energy eigen function. \item The key ingredient that we need for solving the WD equation (\ref{eq21}) is $V(x)$. In case of any modified gravity theory constructed in the background of homogeneous and isotropic space-time, one can get $V(x)$. Further if one can solve the WD equation to find the wave function of the universe and hence its norm $|\Psi|^{2}$ (probability measure on the minisuperspace), this is an important tool for the singularity analysis in the quantum regime. \item If $|\Psi|^{2}=0$ at zero volume ($x^{3}=0$) then singularity is avoided in the sense that probability of having zero volume (singularity) is zero, otherwise the singularity still persists in the quantum description. Therefore an immediate application of this canonical quantization for the present system lies in the singularity analysis of spatially homogeneous and isotropic cosmological models at quantum level.\item Thus the existence (or non existence) of singularity is not a generic one, it depends on the gravity theory under consideration. 
	\end{enumerate}
	\textbf{Remark}: Here $|\Psi|^{2}$ is proportional to the probability measure on the minisuperspace (hence we can normalize it by proper scaling and can apply in the models where the norm is finite). Further the solution of the Wheeler-Dewitt (WD) equation may be interpreted as the propagation amplitude
	of the congruence of geodesics. Norm of this solution (wave function) can be interpreted
	as the probability distribution of the system. If the wave packet so constructed by this
	solution is peaked along the classical solution at the early era then the singularity may
	be avoided so that the geodesics will never converge i.e, if $|\Psi|^{2}\rightarrow0$ as volume of the minisuperspace $\rightarrow0$ then it implies that the quantum description may be able to avoid the initial big-bang singularity (in the sense that probability of the universe (or the minisuperspace model in quantum cosmology) to have zero volume (singularity) is zero). There lies the motivation behind using $|\Psi|^{2}$ as a quantity proportional to the probability density in the present context. Moreover Wheeler-DeWitt quantization scheme is expected to find application in the investigation of the
	singularities in the quantum regime for a collapse of homogeneous systems, such as the
	Datt-Oppenheimer-Snyder collapse.\\ \\
	\textbf{WKB Approximation}\\
	The WKB approximation is a process of transition from quantum solutions to the classical regime. Here the wave function is written as
	\begin{equation}\label{eq23}
		\Psi=\exp\left(\frac{i}{\hbar}S\right)
	\end{equation}with power series expansion in $\hbar$ for $S$ i.e,
	\begin{equation}\label{eq24}
		S=S_{0}+\hbar S_{1}+\hbar^{2}S_{2}+...~~~~,
	\end{equation}
	One recovers the classical solution by constructing a wave packet from $S_{0}$ as,
	\begin{equation}
		\Psi=\int{A(\mathbf{k})~\exp\left(\frac{i}{\hbar}~S_{0}\right)}~d\mathbf{k}
	\end{equation}
	with $\mathbf{k}$ , a parameter. Now substituting (\ref{eq23}) (with $S$ from (\ref{eq24})) into the WD equation (\ref{eq21}) and equating power of $\hbar$, one gets
	\begin{equation}\label{eq26}
		\left(\dfrac{dS_{0}}{dx}\right)^{2}=2 x^{2(\frac{1}{n}-1)}V(x)
	\end{equation}
	Thus from (\ref{eq26}),
	\begin{equation}
		S_{0}=\int{\sqrt{2}~x^{(\frac{1}{n}-1)}\sqrt{V(x)}}dx+k_{0}
	\end{equation} where $k_0$ is a constant of integration.
	So a wave can be constructed as
	\begin{equation}
		\Psi(x)=\int{A(\mathbf{k}) ~\exp\left[\frac{i}{\hbar}S_{0}(\mathbf{k},x)\right]}d\mathbf{k}
	\end{equation}
	with $A(\mathbf{k})$, a sharply peaked Gaussian function. Now a constructive interference occurs if $\dfrac{\partial S_{0}(x)}{\partial \mathbf{k}}=0$ which implies a relation between $\mathbf{k}$ and $x$ i.e. $\mathbf{k}=\mathbf{k}(x)$. So the wave function can now be written as
	\begin{eqnarray}
		\Psi(x)=\int A(\mathbf{k}(x))~ \exp \bigg[ \dfrac{i \sqrt{2}}{\hbar}\int x^{(1-\frac{1}{n})}\sqrt{V(x)}dx  +  \nonumber \\
		k_{0} \bigg]  \dfrac{d\mathbf{k}}{dx}dx
	\end{eqnarray}
	\section{Causal interpretation : bohmian trajectories}
	In this section we adopt an alternative interpretation of quantum mechanics to cosmology. In this ontological interpretation of quantum mechanics \cite{Vilenkin:1983xq}, the quantum effects are carried out by a quantum potential and it is applicable to the minisuperspaces by replacing the Schrödinger equation by the Wheeler- Dewitt equation. The quantum trajectories (known as Bohmian trajectories) are the time evolution of the metric and field variables, obeying the quantum Hamilton-Jacobi equation. These Bohmian trajectories are purely classical for large values of the scale factor and quantum effects become dominant for small value of the scale factor. Some typical superposition of the wave functions resolve the initial singularity but in any case these trajectories will not grow to the size of our universe.\\
	In the metric formulation of Einstein gravity there are four constraints : Super momentum constraints or vector constraints and Hamiltonian constraint or scalar constraint. Due to cosmological principle as the space-time is homogeneous and isotropic so vector constraints are identically satisfied and the quantum version of the scalar constraint equation is nothing but the WD equation i.e,
	\begin{equation}\label{eq31}
		\mathcal{H}~[\tilde{q_{\alpha}}(t)~,~\tilde{p^{\alpha}}(t)]~\Psi(q_{\alpha})=0
	\end{equation}
	where, $p^{\alpha}(t)$ and $q_{\alpha}(t)$ represent the homogeneous degree of freedom obtained from the three metric $q_{ij}$ and the conjugate momenta $\Pi^{ij}$. Now, similar to WKB ansatz one may write,
	\begin{equation}\label{eq32}
		\Psi(q_{\alpha})=R~(q_{\alpha})~\exp\left[\frac{i}{\hbar}~S(q_{\alpha})\right]
	\end{equation}
	Now using (\ref{eq32}) in the WD equation (\ref{eq31}) one gets the Hamilton-Jacobi (H-J) equation
	\begin{equation}\label{eq33}
		\dfrac{1}{2}h_{\alpha\beta}(q_{\rho})\dfrac{\partial S}{\partial q_{\alpha}}\dfrac{\partial S}{\partial q_{\beta}}+ U(q_{\rho})+W(q_{\rho})=0
	\end{equation} 
	where, 
	\begin{equation}
		W({q_{\rho}})=-\dfrac{1}{R}~h_{\alpha\beta}~\dfrac{\partial^{2}R}{\partial q_{\alpha} \partial q_{\beta}}
	\end{equation}
	is termed as the quantum potential, $h_{\alpha\beta}$ denotes the reduction of the supermetric to the given minisuperspace \cite{Chakraborty:2001za} and $U(q_{\rho})$ is the particularization of the scalar curvature density ($-q^{\frac{1}{2}}R)$ of the space-like hyper-surfaces. It may be noted that, due to causal interpretation, the trajectories $q_{\alpha}(t)$ in quantum cosmology must be real and observer independent and the H-J equation will classify them as follows.\\
	The momentum corresponding to $q_{\alpha}$ can be obtained from the above H-J equation (\ref{eq33}) as
	\begin{equation}
		p^{\alpha}=\dfrac{\partial S}{\partial q_{\alpha}}.
	\end{equation}
	Now comparing the above momentum with the usual momentum-velocity relation (i.e $p^{\alpha}~=~f^{\alpha\beta}~\dfrac{\partial q_{\beta}}{\partial t}$) one obtains the quantum trajectories as
	\begin{equation}
		p^{\alpha}=\dfrac{\partial S}{\partial q_{\alpha}}= f^{\alpha\beta}\dfrac{\partial  q_{\beta}}{\partial t}
	\end{equation}
	These first order differential equations are also known as Bohmian trajectories and are invariant under time reparametrization \cite{Chakraborty:2001za}. Hence there will be no problem of time for causal interpretation of minisuperspace quantum cosmology.\\
	Now to obtain the Bohmian trajectories for the present quantum system  the ansatz for the wave function in eq. (\ref{eq32}) reduces to
	\begin{equation}\label{eq90*}
		\Psi(x)= R(x)~\exp\left(\frac{i}{\hbar}~S(x)\right).
	\end{equation}
	Using this ansatz into the WD- equation (\ref{eq21}) one gets the Hamilton-Jacobi equation as 
	\begin{equation}\label{eq91}
		\dfrac{-1}{2~x^{2(\frac{1}{n}-1)}}~\left(\dfrac{dS}{dx}\right)^{2} + W(x) + V(x)=0,
	\end{equation}
	where $W(x)$, the quantum potential has the expression as
	\begin{equation}
		W(x)=\dfrac{1}{2 R(x) x^{2(\frac{1}{n}-1)}} \dfrac{d^{2}R(x)}{dx^{2}}.
	\end{equation}
	Thus the Hamilton-Jacobi function $S$ is given by 
	\begin{equation}
		S=s_{0} \pm \int\left(\dfrac{1}{R(x)}\dfrac{d^{2}R(x)}{dx^{2}}+ 2x^{2(\frac{1}{n}-1)}\right)^{\frac{1}{2}}~dx.
	\end{equation} $s_{0}$ is the constant of integration.\\
	It may be noted that the trajectories $x(t)$ due to causal interpretation should be real, independent of any observation and are classified by the above H-J equation. In fact, the quantum trajectories i.e the Bohmian trajectories are first order differential equations characterized by the equivalence of the usual definition of momentum with that from the Hamilton-Jacobi function $S$ as 
	\begin{equation}
		\dfrac{dS(x)}{dx}=-2 x^{2(\frac{1}{n}-1)} x^{'}
	\end{equation}i.e
	\begin{equation}
		2x^{2(\frac{1}{n}-1)}x^{'}=\mp \left(\dfrac{1}{R(x)}\dfrac{d^{2}R(x)}{dx^{2}}+2x^{2(\frac{1}{n}-1)}\right)^{\frac{1}{2}}
	\end{equation} or,
	\begin{equation}\label{eq96}
		2~\int \dfrac{x^{2(\frac{1}{n}-1)}~dx}{\left(\dfrac{1}{R(x)}~\dfrac{d^{2}R(x)}{dx^{2}}+2x^{2(\frac{1}{n}-1)}\right)^{\frac{1}{2}}}=\mp (t-t_{0}).
	\end{equation} Now we construct the trajectories first without quantum potential and then with quantum potential as follows:\\
	\textbf{\underline{Case-I}}\\
	$R(x)=R_{0}$, a constant then one has,
	\begin{equation}
		S=s_{0} \pm \sqrt{2}~\int x^{(\frac{1}{n}-1)}~dx,
	\end{equation} or
	\begin{equation}
		S=s_{0}\pm \sqrt{2}~n~x^{\frac{1}{n}}
	\end{equation} and the quantum trajectory is described as,
	\begin{equation}
		\sqrt{2}~n~x^{\frac{1}{n}}=\pm (t-t_{0}).
	\end{equation}
	Here the quantum potential is zero and the H-J equation (\ref{eq91}) coincides with the classical one. Thus Bohmian trajectory corresponds to classical power law form of expansion and it can't avoid the initial big-bang singularity. \\
	\textbf{\underline{Case-II}}\\
	Now we study the nature of the trajectories for non-zero quantum potential. Therefore, one may choose $R(x)=x^{N}$, $N\neq0, 1$ (for this choice $W(x)\neq0$) and substituting this in equation (\ref{eq96}) one gets the quantum trajectories as,
	\begin{equation}
		x^{\frac{2}{n}}=\dfrac{(t-t_{0})^{2}}{2n^{2}}-\dfrac{N(N-1)}{2}.
	\end{equation}
	Hence for $N\in (0,1)$ volume is non-zero as $t\rightarrow t_{0}$. Therefore initial big-bang singularity is avoided with non-zero quantum potential and proper fractional power law choice of $R(x)$.
	\section{BRIEF DISCUSSION and conclusion}
	The paper deals with RE both from classical and quantum aspects. The issue of quantization of the RE is a bit tricky if it is treated as a geometric identity in space-time. However in relation to some gravity theory the curvature scalar can be expressed in terms of the energy momentum tensor (and/ or effective energy momentum tensor, in modified gravity theory). In that case Lagrangian formulation due to the Helmholtz conditions has some definite meaning. Further it is generally speculated that $x$ (defined in eq.(\ref{eq1})) is a function of proper time $\tau$ (for time-like geodesic congruence) or cosmic time $t$ (in cosmology)  and so its derivatives. Therefore it is reasonable to consider $x$ and its derivative as functionally related. As a consequence the functional $V[x]$ becomes simply a function $V(x)$.\\
	$~~~~~~~~~~~~$ Moreover there is another issue related to quantization of RE, applicable to cosmological models, particularly for spatially an-isotropic space-time. In canonical quantization scheme the evolution will be non-unitary in nature. However this may be resolved (i.e. evolution may be made unitary) by suitable choice of the operator ordering. However for homogeneous models it is always possible to have a self adjoint extension using the result of Friedrichs \cite{Pal:2016ysz}. Hence it is always expected to have an unitary evolution of geodesic congruence due to self-adjoint extension of the Hamiltonian.\\ In order to have a Lagrangian and Hamiltonian formulation, the RE has been transformed into a second order nonlinear differential equation by a suitable transformation related to the metric of the hyper-surface. A first integral of this transformed second order differential equation may be considered as a solution to the RE. As the RE can be written as a second order differential equation so it has been shown to have a general formulation of Lagrangian and Hamiltonian for it. A canonical quantization scheme with construction of WD equation and WKB approximation has been carried out. Finally, Bohmian trajectories have been derived by constructing the Hamilton-Jacobi equation with quantum potential and applied to the present quantum system. Here the quantum Bohmian trajectories unlike classical geodesics are able to obviate the initial big-bang singularity in the presence of non-zero quantum potential and with proper power-law choice of the pre factor present in the ansatz for the wave function. \\$~~~~~~~$So this paper aims at showing two possible pathways of avoiding singularity quantum mechanically: (i) By Canonical quantization method where the wave function (most specifically norm of the wave function) is determined by solving the WD equation for spatially homogeneous and isotropic cosmological models for the sake of singularity analysis and, (ii) Bohiman formulation (independent of any gravity theory) where suitable choice of the wave function of the universe helps to avoid the classical singularity in the presence of quantum potential. 
	\section*{Acknowledgement}
	The authors are thankful to the anonymous reviewers for their insightful comments on the manuscript which improved the quality of the paper. M.C thanks University Grants Commission (UGC) for providing the Junior Research Fellowship (ID:211610035684/JOINT CSIR-UGC NET JUNE-2021). S.C. thanks FIST program of DST, Department of Mathematics, JU (SR/FST/MS-II/2021/101(C)).


\begin{thebibliography}{50}
		\bibitem{Raychaudhuri:1953yv}
		A.~Raychaudhuri,
		``Relativistic cosmology. 1.,''
		Phys. Rev. \textbf{98}, 1123-1126 (1955).
		\bibitem{Dadhich:2007pi}
		N.~Dadhich,
		Pramana \textbf{69}, 23-30 (2007)
		\bibitem{Dadhich:2005qr}
		N.~Dadhich,
		``Derivation of the Raychaudhuri equation,''18 Nov (2022)[arXiv:gr-qc/0511123 [gr-qc]].
		\bibitem{Ehlers:2006aa}
		J.~Ehlers,
		Int. J. Mod. Phys. D \textbf{15}, 1573-1580 (2006).
		\bibitem{Kar:2008zz}
		S.~Kar,
		Resonance J. Sci. Educ. \textbf{13}, 319-333 (2008).
		\bibitem{Kar:2006ms}
		S.~Kar and S.~SenGupta,
		Pramana \textbf{69}, 49 (2007).
		\bibitem{Penrose:1964wq}
		R.~Penrose,
		Phys. Rev. Lett. \textbf{14}, 57-59 (1965).
		\bibitem{Hawking:1970zqf}
		S.~W.~Hawking and R.~Penrose,
		Proc. Roy. Soc. Lond. A \textbf{314}, 529-548 (1970).
		\bibitem{Hawking:1973uf}
		S.~W.~Hawking and G.~F.~R.~Ellis,
		Cambridge University Press, (2011).
		\bibitem{Thiemann:2007pyv}
		T.~Thiemann,
		Cambridge University Press, (2007)
		\bibitem{DeWitt:1967yk}
		B.~S.~DeWitt,
		Phys. Rev. \textbf{160}, 1113-1148 (1967)
		\bibitem{Hawking:1978jz}
		S.~W.~Hawking,
		Phys. Rev. D \textbf{18}, 1747-1753 (1978)
		\bibitem{Hartle:1983ai}
		J.~B.~Hartle and S.~W.~Hawking,
		Phys. Rev. D \textbf{28}, 2960-2975 (1983)
		\bibitem{Hartle:2022ykc}
		J.~B.~Hartle,
		``Simplicial Quantum Gravity,''
		[arXiv:2201.00226 [gr-qc]].
		\bibitem{Vilenkin:1988yd}
		A.~Vilenkin,
		Phys. Rev. D \textbf{39}, 1116 (1989)
		\bibitem{Das:2013oda}
		S.~Das,
		Phys. Rev. D \textbf{89}, no.8, 084068 (2014)
		\bibitem{Ali:2014qla}
		A.~F.~Ali and S.~Das,
		Phys. Lett. B \textbf{741}, 276-279 (2015)
		\bibitem{Blanchette:2021vid}
		K.~Blanchette, S.~Das and S.~Rastgoo,
		JHEP \textbf{09}, 062 (2021)
		\bibitem{Burger:2018hpz}
		D.~J.~Burger, N.~Moynihan, S.~Das, S.~Shajidul Haque and B.~Underwood,
		Phys. Rev. D \textbf{98}, no.2, 024006 (2018)
		\bibitem{Blanchette:2020kkk}
		K.~Blanchette, S.~Das, S.~Hergott and S.~Rastgoo,
		Phys. Rev. D \textbf{103}, no.8, 084038 (2021)
		\bibitem{Alsaleh:2017ozf}
		S.~Alsaleh, L.~Alasfar, M.~Faizal and A.~F.~Ali,
		Int. J. Mod. Phys. A \textbf{33}, no.10, 1850052 (2018)
		\bibitem{Davis:1928}
		D.~R.~Davis,
		Trans. Amer. Math. Soc. \textbf{30} (1928), 710-736
		
		\bibitem{Davis:1929}
		D.~R.~Davis,
		Bull. Amer. Math. Soc. \textbf{35} (1929), 371-380
		
		\bibitem{Douglas:1941}
		J.~Douglas,
		Trans. Amer. Math. Soc. \textbf{50} (1941), 71-128
		
		\bibitem{Casetta:1941}
		L.~Casetta, C.~P.~Pesce
		Trans. Amer. Math. Soc. \textbf{50} (1941), 71-128
		
		\bibitem{Crampin:2010}
		M.~ Crampin, T.~ Mestdag and W.~ Sarlet
		Z. Angew. Math. Mech. \textbf{90} (2010), 502-508
		
		\bibitem{Nigam:2016}
		K.~ Nigam, K.~ Banerjee
		``A Brief Review of Helmholtz Conditions" 
		\bibitem{Vilenkin:1983xq}
		A.~Vilenkin,
		Phys. Rev. D \textbf{27}, 2848 (1983)
		\bibitem{Pal:2016ysz}
		S.~Pal and N.~Banerjee,
		J. Math. Phys. \textbf{57}, no.12, 122502 (2016)
		\bibitem{Chakraborty:2001za}
		S.~Chakraborty,
		Int. J. Mod. Phys. D \textbf{10}, 943-956 (2001)
	\end{thebibliography}
\end{document}